\documentclass{agnspec}
\usepackage{graphics}
\usepackage{psfig}
\def\R{~ROSAT}

\def\lesssim{\mathrel{\hbox{\rlap{\hbox{\lower4pt\hbox{$\sim$}}}\hbox{$<$}}}}
\def\gtrsim{\mathrel{\hbox{\rlap{\hbox{\lower4pt\hbox{$\sim$}}}\hbox{$>$}}}}
\def\araa{Ann. Revs. Astron. Astrop. }
\def\apj{Ap.J. }
\def\apjl{Ap.J. Letts. }
\def\mnras{M.N.R.A.S. }
\def\pasj{P.A.S.J. }
\def\aap{Astron. Astrop.}
\def\kms{km~s$^{-1}$}

\begin{document}

\title{High Resolution X-ray Spectroscopy and the Nature of Warm Absorbers
in AGN}

%\titlerunning{Multi-Temperature Warm Absorbers}

\author{Julian H. Krolik}
\institute{Department of Physics and Astronomy, Johns Hopkins
University, Baltimore MD 21218 USA}

\maketitle

\begin{abstract}

   Although soft X-ray absorption features in AGN were discovered almost
ten years ago, the nature and location of the gas creating them has remained
controversial.  However, by making use of the newly-available high-resolution
spectra provided by {\it XMM-Newton} and {\it Chandra}, we should be able
to make substantial advances.  The first such spectra indicate that multiple
ionization states coexist in the absorber; this is a natural consequence
of photoionization physics.  Photoionized
evaporation in the presence of a copious mass source locks the ratio
of ionizing intensity to pressure to a critical value.
A broad range of temperatures can all coexist in equilibrium for
this value of the ratio of ionizing intensity to pressure.  Consequently,
the flow is expected to be strongly inhomogeneous in temperature.
The inferred distance of this material from the source of ionizing
radiation depends on how much matter exists at the highest-obtainable
temperature.  This distance can be measured by monitoring how ionic
column densities respond to changes in the ionizing continuum on
timescales of days to years.
\end{abstract}

\section{The Unanswered Questions about Warm Absorbers: How {\it Chandra}
and {\it XMM} Can Help}

    Although absorption features due to highly-ionized species were first
clearly seen in AGN spectra taken with \R\ almost ten years ago (Turner et al.
1993), most of the fundamental facts about them remain obscure today.  This is
true at even the most basic, empirical level.

\subsection{The ionization distribution}

    For example, we do not even know which ionization states are present in
the absorbing gas.  Because increasing levels of
ionization tend to move atomic features higher in energy and then remove
them altogether, the character of ionization states possible to see in
X-ray spectra is strongly influenced by the specifics of instrumentation.
Consequently, our efforts to determine how much matter there is in
which ionization state suffer from extremely strong instrumental
selection effects.

   In \R-data, the spectral resolution and signal/noise were so poor
that the only feature strong enough to be seen was the O
K-edge, and the ionization state of O could be determined only roughly.
Labelling ionization level by the ionization parameter $\xi \equiv
L_{ion}/nr^2$ (where $L_{ion}$ is conventionally defined as the
luminosity between 1 and 1000 Ryd, $n$ is the H-nucleus density, and
$r$ is the distance to the source of radiation), this sort of data
was capable only of identifying the presence of matter with $\xi \lesssim 30$,
and perhaps distinguishing ionization parameters at the highest end of
this range from those an order of magnitude or more smaller.

   Unfortunately, \R\ was was totally incapable of detecting any matter
more strongly ionized because higer ionization destroys OVIII.
Other elements' K-edges could not be seen in \R\ data either because their
energy was outside its limited range (e.g., Fe) or because their
abundance was too  small to permit detection with such crude data.
To put this (semi)-quantitatively, in gas with solar abundances most
of whose O has not been stripped, O K-edges were detectable by \R
when the column density $N \gtrsim 10^{21}$~cm$^{-2}$.  Even if the
\R\ energy range extended to the K-edge of ionized Fe at 7--9~keV,
a column density almost three orders of magnitude larger would have
been necessary in order for a significant edge to have been formed.
In this sense, \R-level data was roughly 1000 times less sensitive
to highly-ionized than to weakly-ionized gas.

    Data of the sort produced by ASCA was better, but not greatly.  Although
its resolution and signal/noise were good enough to permit estimates of
the depths of the OVII and OVIII edges separately (e.g., Reynolds et al. 1995),
it could see no other edge features reliably.  The next-deepest K-edges
(Si, S, Ar, etc.) are so much smaller in opacity that they would be
detectable in ASCA data only for much greater absorbing columns.
In principle, ASCA's nominal energy range included the Fe K-edge, but
the effective area there was so small as to make searching for this edge
impractical.  Thus, like \R, ASCA could
detect matter with $\xi \lesssim 30$, but could improve upon \R\ by
permitting $\xi \simeq 30$ to be distinguished from, say, $\xi = 3$.  Again
like \R, it was effectively blind to more highly-ionized matter.

    The advent of the grating spectrometers on {\it Chandra} and {\it XMM}
has already dramatically changed this frustrating situation, and will
undoubtedly improve matters further in the near future.  Their tremendously
finer spectral resolution permits study of line transitions, not just
edges, while their much greater signal/noise allows one to search for
much weaker features.   With data of this quality, it is possible to
search for species expected to be common
up to $\xi \sim 1000$ even when the column density of that gas is as
small as $\lesssim 10^{21}$~cm$^{-2}$---and we are starting to find them.
Several different recent warm absorber studies employing grating data
report that more than one ionization component is detected, with
$\xi$ ranging anywhere from a few tens to $\sim 1000$ (Kaastra et al.
2000, Collinge et al. 2001, Kaspi et al. 2001).

    Thus, in the next few years we may reasonably hope to make much
progress on the most basic empirical question of warm absorber studies:
measuring the distribution $dN/d\xi$.

\subsection{Location and origin}

   After identifying the absorber's ionization state, the next obvious
question is where to place it in the AGN system.  Proposed distances have
ranged all the way from inside or near the broad-line region at 0.01 -- 0.1~pc
from the nucleus (Reynolds et al. 1995, George et al. 1998, Netzer et al. 2001)
to many tens of pc away in the narrow-line region (Bottorff et al. 2000).
Some have argued that there is absorbing material across this entire range
of radii (Otani et al. 1996, Morales et al. 2000).  Others (e.g., Krolik
\& Kriss 1995) have argued that the absorbing gas should be identified with the scattering
region posited by Seyfert galaxy unification schemes, perhaps $\sim 1$~pc
from the nucleus when $L_{ion} \sim 10^{44}$~erg~s$^{-1}$.

   The state of affairs regarding the absorbing matter's origin is even worse.
A few of the diverse suggestions that have been floated include: evaporation
off ``bloated stars" in the broad-line region (Netzer 1996); gas evaporated
off the torus obscuring the nucleus (Krolik \& Kriss 1995); and a wind
driven off the accretion disk (Elvis 2000; Bottorff et al. 2000).  The
absorbing gas's ultimate fate has been left wholly unconsidered.

    This, too, is a topic on which we may hope for enlightenment from new
spectroscopy data.  As has been remarked in numerous papers (Reynolds
et al. 1995, Krolik \& Kriss 2001, Netzer et al. 2001), variability
in absorption features can strongly constrain the location of the
gas responsible.  The virtue of the new data sets is that, by making
so many features observable, our odds of seeing variability in one or
more of them are greatly improved.  Moreover, because the new data are
less subject to ionization selection effects (as discussed in the
previous subsection), the location bounds placed by variability can
be applied more generally.

\subsection{Kinematics}

    Independent of where the absorbing gas is located, we'd also like to
know which way it's moving.  For this issue, high-resolution spectroscopy is
essential.  If the absorbing gas possesses ions with ionization potentials
$\sim 1$~keV (as is necessary to have numerous soft X-ray features), its
temperature cannot be much greater than $\sim 10^7$~K.  Transonic motions
at this temperature would then be $\sim 300$~km~s$^{-1}$ or slower.  It
immediately follows that spectral resolution $\Delta E/E \gtrsim 1000$
is a prerequisite for beginning to detect motions at this speed.

    Because resolution of this caliber is exactly what {\it Chandra}
and {\it XMM} have now delivered, we are beginning to learn about this
question.  Kaspi et al. (2001), for example, have shown that the absorber
in NGC~3783 is moving toward us (relative to the nucleus) at
$\simeq 500$~km~s$^{-1}$.

\subsection{Geometry}

     Although the depth of an edge may tell us the total column density of
ions along the line of sight, it does not tell us whether they are all
clumped together, or spread out smoothly, or clustered at a variety of
locations.  And, of course, absorption tells us nothing about the existence
of matter along other directions.

     Other constraints are necessary to make progress answering questions
of this sort.  For example, resonance line emission components offset toward
the red from absorption components are the classic signature of outflows.
The relative equivalent width of the emission and absorption components
reveals the ratio between the mean optical depth on lines of sight other
than ours to the ratio on the line toward us.  Similarly, the relative
velocity widths further constrain the gas's geometric disposition.
Making use of the emission/absorption equivalent width indicator,
Kaspi et al. (2001) have already suggested that the absorbing gas in
NGC ~3783, while occupying a large part of solid angle around the
nucleus, does not completely cover it.

     Clumpiness in the gas can be analyzed in any of several ways, given
quality spectroscopy:  Emission lines that are either collisionally-excited or
created by recombination yield measurements of the gas's emission measure;
combined with independent density estimates, one can then infer the total
emitting volume and compare it to the available volume in the region.
Alternatively, spectroscopic data that requires gas at several
different ionization levels co-existing in more or less the same location
would strongly suggest a clumpy, inhomogeneous environment.

\subsection{Dynamics}

    Given our ignorance about the warm absorber's kinematics, location, and
origin, it should come as no surprise that almost nothing is known with
confidence about its dynamics.  We have as yet advanced hardly at all past
listing the ``usual suspects": gravity, thermal presure gradients (Balsara
\& Krolik 1993), radiation pressure (Chelouche \& Netzer 2001, Morales
\& Fabian 2002), and magnetic fields (Bottorff et al. 2000).

\section{Warm Photoionized Equilibria}

    To make further progress, it is helpful to consider a few constraints
imposed by the basic facts of photoionization physics.  We begin with the
most elementary of these, a relationship that links the ionization parameter,
the distance, the gas column density, and its volume filling factor.  From
the definition of $\xi$ and the fact that the column density $N = n\Delta r$,
where $\Delta r$ is the total distance along the line of sight occupied by
the absorbing gas, we immediately see (as shown by Turner et al. 1993) that
\begin{equation}  
   r = {L_{ion} \over N \xi} {\Delta r \over r}.  
\end{equation}

Because $\Delta r/r$ cannot possibly be greater than unity, the inferred
ionization parameter and column density of an absorber, when combined with
the luminosity of the ionizing source, place an upper bound on its distance
from that source:
\begin{equation}
    r \leq r_{max} \equiv {L_{ion} \over N \xi}.   
\end{equation}
For typical numbers, $r_{max} \sim 30 L_{ion,44}N_{22}^{-1}(\xi/100)^{-1}$~pc.
Here we have scaled the ionizing luminosity to $10^{44}$~erg~s$^{-1}$ and
the column density to $10^{22}$~cm$^{-2}$.  For fixed $r_{max}$, the more
highly clumped the gas is, the smaller the radius at which it might be found.

     A second constraint is provided by the nature of photoionization
equilibrium.  If the gas's cooling time is shorter than the time required
for it flow through the absorbing region, we can expect that it is in
energy balance with the radiation.  If that is the case, its temperature
is a unique function of its ionization parameter $\xi$.

     If we wish to link photoionization physics with fluid dynamics, it
is more convenient to write the ionization parameter in a related, but
significantly different form: $\Xi \equiv \xi /(4\pi c kT)$.  The factor
of $4\pi c k$ makes $\Xi$ renders it dimensionless; more significantly,
by dividing by the temperature, a given $\Xi$ at any particular radius
corresponds to a certain pressure.  Viewed from the opposite perspective,
if the boundary condition is a fixed pressure and the gas is in energy
balance, $\Xi$ is determined, but $T$ is uniquely determined only when
$T(\Xi)$ is single-valued.  Although $T(\xi)$ is, in general, unique,
when the ionizing continuum has a form typical of AGN, $T(\Xi)$, in
general, is {\it multi-valued} in the range of temperatures between
$\simeq 5 \times 10^4$~K and $\gtrsim 10^6$~K (Krolik, McKee \& Tarter 1981;
fig.~1).
As a result, the boundary conditions may not suffice to determine the
temperature; its history (and possibly the influence of thermal instabilities)
is instead the final deciding factor.   The fact that the thermal balance
may be multi-valued is central to the study of warm absorbers because the
multiple-valued solutions occur in exactly the range of conditions capable
of making soft X-ray features.

\begin{figure}
\centerline{\psfig{figure=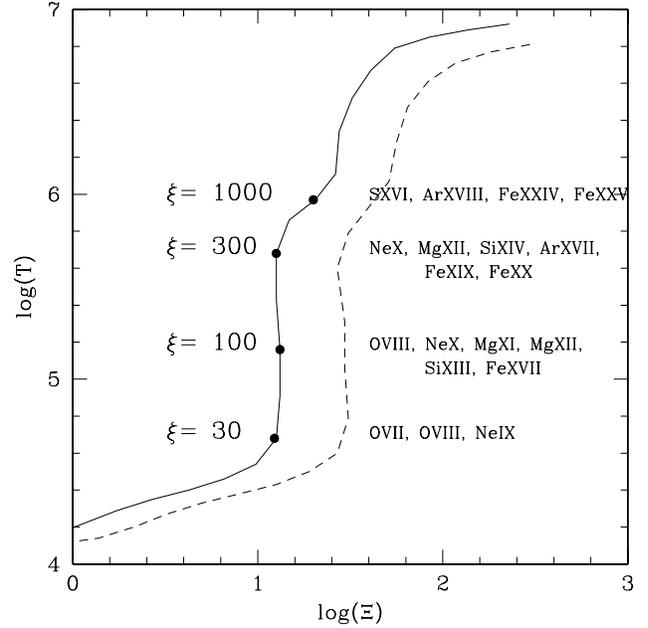,width=8.8cm}}
\caption[]{The temperature equilibrium curve for two guesses (Krolik \& Kriss
2001--solid curve, Kaspi et al. 2001--dashed curve) about the unknown
shape of the EUV spectrum in Seyfert galaxies.  Although the value
of $\Xi$ at which the nearly vertical rise between $5 \times 10^4$~K and
$1 \times 10^6$~K takes place depends on the shape of the spectrum, the
character of the rise is a persistent feature of all reasonable AGN
spectra.   The location of several values of $\xi$ is shown, as are a
few of the dominant ionization stages associated with each of the marked
values.}
\end{figure}

     When a detailed calculation is made of the thermal equilibrium curve,
it is necessary to make several guesses about matters that we don't truly
know and yet have noticeable effect on the result.   The distribution of
elemental abundances is the first of these.  Generally-speaking, solar
abundances are assumed; other choices can alter the detailed shape of the
curve, but within quite wide bounds have at most modest effect on the
shape of $T(\Xi)$.  More important to the thermal balance curve is the
shape of the ionizing continuum.  In low-redshift objects, we can observe
the continuum at energies just below 1 Rydberg and again above
$\simeq 500$~eV, but the most important segment of the ionizing continuum
is precisely in the band from 13~eV to $\simeq 500$~eV that we can never
directly see.  The best we can do is smoothly interpolate from either
side.   Fortunately, as shown in Figure~1, different plausible guesses
do not result in qualitatively different results.  Both equilibrium
calculations shown in the figure were based on continuum shapes meant
to apply to the Seyfert galaxy NGC 3783 (one in Kaspi et al. 2001,
the other in Krolik \& Kriss 2001).  Although rather different, both
predict the same general sort of behavior---a nearly vertical rise in
temperature from $\simeq 5 \times 10^4$~K to almost $10^6$~K.  The
value of $\Xi$ where this takes place is called $\Xi_c$.

   In fact, it is worthwhile to focus greater attention on the
``vertical branch" of the photoionized thermal equilibrium curve because
it is significant for several reasons.  The first, as just mentioned, is
that the range of $\xi$ inferred from the ionization stages seen in
warm absorber spectra is $\sim 10$ -- 1000; this is almost exactly the range
traversed on the vertical branch.  In other words, if warm absorbers
are in a state of photoionization equilibrium and thermal balance, they
``live" on the vertical branch.

   Second, the vertical branch has a rather special thermodynamic property:
it is marginally stable to isobaric thermal perturbations.  Were
$dT/d\Xi > 0$, constant pressure perturbations would be stable because
increasing the temperature from an initial point on the equilibrium curve
must result in net cooling; conversely, were $dT/d\Xi < 0$, the same sort
of perturbation grows exponentially because the excursion causes net heating.
However, on the vertical branch $d\Xi/dT = 0$: in this case, if
the temperature is perturbed while the pressure is fixed, the gas remains
in a state of zero net radiative energy gain or loss.  Consequently,
there is no temperature ``feedback", either positive or negative.
Perturbations are neither strongly damped nor strongly amplified---they simply
do whatever the outside agency creating them demands.   Although it is true
that the curve of thermal balance can never be precisely vertical, to the
extent that it is nearly so, the net cooling or heating that is
engendered by the perturbations is small, and the magnitude of growth
permitted in a region of weak instability is capped at a low level
as a new segment of the equilibrium curve is quickly reached.

    Third, far from being a special case, the value of $\Xi$ at which
the vertical branch occurs is
actually a preferred value of the pressure-based ionization parameter.
Whenever large amounts of cool material are exposed to a strong ionizing
flux (as might readibly be imagined near an AGN), the low-density
portion of this cool matter finds itself rapidly ionized and heated.
If the heating time is short compared to the flow time, the ambient
pressure rises equally rapidly.  At fixed distance from the source
of ionizing radiation, this means the value of $\Xi$ surrounding the
cool matter rapidly falls.  Nothing stops this fall in $\Xi$ until it
reaches $\Xi_c$, where the net energy exchange via radiation switches
from heating to cooling.   Similarly, if the
pressure overshoots to the net-cooling side of the vertical branch,
matter cools rapidly, losing pressure, until balance is achieved---at
exactly $\Xi_c$.
    
    This portion of the argument may be summed up very simply: Given that,
by definition, warm absorbers are exposed to AGN fluxes, it makes sense
to suppose that they are photoionized.  Let us suppose for the moment
that the gas is in photoionization equilibrium and radiative balance.
Then the particular ions observed (OVII,VIII; NeX; MgXII; SiXIV; etc.)
indicate that the gas lies on the vertical branch of the $T(\Xi)$ curve.
Because that branch allows a wide range of temperatures to coexist at
a single pressure, it follows that the material is likely to
contain multiple sub-regions whose temperature could be
anywhere within the range found on the vertical branch
of the equilibrium curve.  Moreover, pressure balance tends to force
photoionized gases in the presence of a copious mass source toward
$\Xi_c$.  That is, it is a natural consequence of photoionization
physics that we should see gas with $\Xi = \Xi_c$ and that it should
be inhomogeneous in temperature.

\section{Global Picture}

     These special thermodynamic properties of the vertical branch find a
natural application in the context of AGN warm absorbers.  These are
frequently seen in type 1 Seyfert galaxies, a variety of AGN in which
we have excellent reason to believe that extremely optically thick
obscuring matter wraps around the nucleus with roughly toroidal geometry
(Antonucci 1993, Krolik 1999).
Where the ionizing radiation of the nucleus strikes the inner edge of
the obscuring stuff, exactly the process described in the previous
paragraph can be expected to take place: copious amounts of matter should
be ionized and heated, maintaining $\Xi \simeq \Xi_c$.   Until the temperature
rises quite high, the heating time is far shorter than the flow time: as shown
in detail in Krolik \& Kriss (2001), if the flow speed is of the magnitude
observed ($\simeq 500$~\kms), this criterion is met for $T < 4 \times 10^6
r_{pc}^{-1/2}$~K, for $r_{pc}$ the distance from the central source in parsecs.

      As the newly-liberated material is heated, it will expand in order
to maintain pressure balance.  One might then ask, ``Will the expanding
parcels eventually be confined?  Or will they spread out and fill the
entire volume of the region?"   Suppose that the matter we
see by its ``warm absorber" features is clumpy.  If the volume outside
the clumps has lower pressure, the clumps will expand at roughly their
internal sound speeds.  The ratio in column density between the low-pressure
background and the high-pressure absorbing lumps may then be estimated as
\begin{equation}
N_{bkgd}/N_{abs} \sim (r/\Delta r_{abs})(c_s/v_{flow}),
\end{equation}
where the absorber has radial thickness $\Delta r_{abs}$ and sound speed
$c_s$.  Because the sound speed at $\sim 10^5$~K (the typical temperature
inferred for warm absorber regions) is $\sim 50$~\kms, it is clear that
the column density of the background is {\it automatically} close to the
column density of the absorber if there is strong clumping.  That is, one
must expect the background to be a significant part of the absorbing
structure.  Put yet another way, it is probably best to consider the entire
system as volume-filling, albeit one that is likely to be strongly
inhomogeneous.

     Although the ionization parameter $\Xi$ is tightly constrained by
photoionized thermodynamics, the temperature, at least within the range
permitted at $\Xi \simeq \Xi_c$, is hardly constrained at all.  We should
expect, then, that the evaporating matter in this region might exist
anywhere in the temperature range $5 \times 10^4$ -- $\sim 10^6$~K,
i.e., $30 \lesssim \xi \lesssim 1000$.  The issue is not ``What is the
favored value of $T$ or $\xi$?", but ``What is the distribution of $T$
and $\xi$ within this broad range?"  Small accidents having to do with
irregularities in the source of matter (the obscuring torus?) or ``bumps"
in the outflow of heated gas could affect this distribution, so different
objects could easily differ in detail.

     At fixed pressure, the hottest regions have the least density and so,
per unit mass, occupy the largest volume.  For this reason, one might
expect most of the volume to be occupied by relatively hot gas, but it
is likely studded by smaller regions of higher density and lower temperature.
Each of these sub-regions can easily change its temperature over time.

\section{Implications}

     This conceptual picture has numerous implications for the character of
warm absorbers.  The most basic, of course, is that the features we see
should originate in ions found in photoionized equilibrium over a span of a
factor of 20 or so in temperature.  There are also other, more quantitative
implications.
      
\subsection{Location}

      Returning to equations~1 and 2, we can rewrite them as
\begin{equation}
r \simeq r_{max}(\xi_{max}) = {L_{ion} \over \xi_{max} N(\xi_{max})},
\end{equation}
where $\xi_{max}$ is the highest $\xi$ (corresponding to the highest
$T$) attainable in a flow time.  Because the gas at $\xi_{max}$ is the
hottest that we can expect to find in the absorber region, it should
occupy the largest part of the volume.  That is, for the hottest
gas $\Delta r/r \simeq 1$, while cooler gas is strongly clumped.
Scaled in terms of typical numbers, the predicted distance
\begin{equation}
r \simeq 3 {L_{ion,44} \over N_{22}}\hbox{ pc},
\end{equation}
where $\xi_{max}$ has been set at 1000.

      A few parsecs is, of course, the scale of the inner torus (as seen,
for example, in near-infrared imaging---Thatte et al. 1997, Marco \& Alloin
2000---or H$_2$O maser spots---Greenhill et al. 1996).  It is likewise
the expected scale of the polarizing reflection region that permits us to
see type 1 Seyfert nuclei in type 2 Seyfert galaxies.

\subsection{Dynamics}

    With an estimate of the distance to the nucleus, the strength of
gravity may be more quantitatively estimated.  Making use of equation~5,
we find that the free-fall speed in the absorber region is
$\simeq 30 (L/L_E)^{-1/2} N_{22}^{1/2}$~\kms.  Here $L_E$ is the Eddington
luminosity and we have supposed that most of the bolometric luminosity is
in the ionizing band.  Because the sound speed in the hottest phase of the
absorber is $\sim 100$~\kms, we can expect that thermal
pressures alone may be able to drive an outflow unless $L/L_E \ll 1$.
Whether radiation or magnetic forces supplement the thermal pressure gradient
remains an open question, but the fact that we see blue-shifted absorption
lines indicating an outflow should now be no surprise.

\subsection{Geometry}

      If the warm absorber is formed by this sort of photoionized evaporation
off the inner edge of the obscuring torus, the torus itself blocks the warm
gas from expanding to the side.  A biconical expansion and outflow results.
For this reason, we might expect the warm absorber to occupy fully only a
part of solid angle around the nucleus.  Moreover, depending on our viewing
angle, parts of the obscuring matter might block our view of selected
portions of the far side of the outflow.  Thus, emission line equivalent
widths should indicate substantial, but incomplete, solid angle coverage---as
is tentatively seen in NGC~3783 (Kaspi et al. 2001).

\subsection{Variability and location}

     This picture also makes specific predictions about variability in
absorption features.  The column densities we see are, of course, controlled
by ionization balance in the line-of-sight gas.  The governing equation
can be written (in abridged form) as
\begin{eqnarray}
{d \ln n_i \over dt} = {n_{i-1} \over n_i}F_x^{\prime}\sigma_{ion}^{\prime}
     \quad + &n_e \alpha_{rec}^{\prime\prime}{n_{i+1} \over n_i} + \ldots \\
     &- F_x \sigma_{ion} - n_e \alpha_{rec} - \ldots
\end{eqnarray}
where the density of ionization stage $i$ is $n_i$, $i\pm1$ correspond to
the stages once more or less ionized, $F_x$ is the photon flux at the
relevant ionization edge, $\sigma_{ion}$ is the appropriately averaged
ionization cross section, and the primes and double primes denote the
quantities relevant
to the $i+1$ and $i-1$ ionization stages.  When $n_{i+1} \gg n_i$ (as is
often true for the particular ionization stages seen), two terms dominate
the rest, so that
\begin{equation}
{d\ln n_i \over dt} \simeq n_e \alpha_{rec}^{\prime\prime} n_{i+1}/n_i -
       F_x \sigma_{ion} .
\end{equation}
The characteristic time for effecting changes in the absorbing column
is then
\begin{equation}
t_{ion} \sim 10^{6 \pm 1} {r_{pc}^2 \over L_{ion,44}}\hbox{ s}.
\end{equation}
Different species have varying ionization timescales, depending on the
details of their atomic structures.  For example, if one examines the
more abundant species in the relevant ionization conditions, SiXIV,
FeXXV, and FeXXVI tend to have longer ionization times, while
OVI and the ions around FeXX tend to have somewhat shorter
timescales.  The range is not terribly large, however: the $\pm 1$ in the
exponent in equation~9 more or less comprises the range for the species of
greatest interest (Krolik \& Kriss 2001).

      Timescales of days to years are, of course, very convenient timescales
for human observations.  Because we know which species should vary on which
timescales (parameterized by the ionizing flux in the absorbing region),
monitoring spectroscopy should be a very powerful tool for constraining
the location and conditions of the absorbing gas.

\subsection{How would the warm absorber look when seen from the side?}

    Finally, it is striking how much the warm absorber region resembles
the reflection region that allows us to see the nuclei of type 2 Seyfert
galaxies.  Both regions appear to lie at distances $\sim$~pc; both are
highly-ionized; both are flowing outward at speeds of several hundred
\kms; the total column densities in both are $\sim 10^{22}$ --
$10^{23}$~cm$^{-2}$.  It would appear to make a great deal of sense
to suppose that they are the same structure, simply viewed at different
angles.

\end{document}